# A sufficiently fast algorithm for finding close to optimal junction trees

**Ann Becker and Dan Geiger**
Computer Science Department
Technion
Haifa 32000, ISRAEL
anyuta@cs.technion.ac.il, dang@cs.technion.ac.il

## Abstract

An algorithm is developed for finding a close to optimal junction tree of a given graph $G$. The algorithm has a worst case complexity $O(c^k n^a)$ where $a$ and $c$ are constants, $n$ is the number of vertices, and $k$ is the size of the largest clique in a junction tree of $G$ in which this size is minimized. The algorithm guarantees that the logarithm of the size of the state space of the heaviest clique in the junction tree produced is less than a constant factor off the optimal value. When $k = O(\log n)$, our algorithm yields a polynomial inference algorithm for Bayesian networks.

## 1    Introduction

All exact inference algorithms for the computation of a posterior probability in general Bayesian networks have two conceptual phases. One phase handles operations on the graphical structure itself and the other performs probabilistic computations; The junction tree algorithm [LS88] requires us to first find a "good" junction tree and then perform probabilistic computations on the junction tree and the method of conditioning [Pe86] requires to find a "good" loop cutset and then perform a calculation using the loop cutset. In [BG94], we offered an algorithm that finds a loop cutset for which the logarithm of the state space is guaranteed to be a constant factor off the optimal value. In this paper, we provide a similar optimization for the junction tree algorithm.

We shall first restrict our discussion to networks for which all vertices have the same state space size and to the optimality criterion which we call cliquewidth. The *cliquewidth* of an undirected graph $G$ is the size of the largest clique in a junction tree of $G$ in which the size of the largest clique is minimized. A more common term is *treewidth* which is the cliquewidth minus 1.

To date all methods in the AI and Statistics communities for finding a junction tree had no guarantee of performance and could perform rather poorly when presented with an appropriate example. One algorithm, due to Rose (1974), is as follows: repeatedly, select a vertex $v$ with minimum number of neighbors $N(v)$, delete $v$ from the graph, and make a clique out of $N(v)$. The resulting sequence of cliques creates a junction tree. This greedy algorithm minimizes the size of each clique as it is being created. However, it could easily make a mistake at the first step that would lead it to a junction tree far off the optimal size. Another algorithm, investigated by Kjaerulff (1990), is simulating annealing which takes a long time to run and has no guarantees on the quality of the output.

Finding an optimal junction tree is NP-complete but for a graph with $n$ vertices and a cliquewidth $k$ there exits an $O(n^{k+1})$ algorithm that finds an optimal junction tree [ACP87]. This algorithm is not practical for the size of Bayesian networks dealt in practice. Other algorithms for finding an optimal junction tree have a complexity of $O(f(k)n)$ where $f(k)$ is a super-exponential function of $k$ [Bo93]. These algorithms are practical for cliques of size $k = 5$ at most. A more practical algorithm for constructing an optimal junction tree when the largest clique size is 4 is given in [AP86]. For larger values of $k$ there is no algorithm to date that can find the optimum junction tree quickly. The exponential dependency in $k$ cannot be improved unless $P = NP$ because finding an optimal clique tree for $k = O(n)$ is NP-complete [ACP87].

Kloks in his book *treewidth* [Kl94], which is devoted to finding junction trees in various graphs, gives a *polynomial* algorithm that finds a junction tree of a given graph $G$ such that its maximal clique size is at most $12\Delta \log n$ off optimal where $\Delta$ is a large unspecified constant (See also, [BGHK91]). Kloks states that finding a polynomial algorithm that constructs a junction tree such that its maximal clique is a constant factor off optimal is a major open problem. The importance of this problem stems from the fact that many NP-complete problems on graphs can be solved polynomially if the input graph has a junction tree with fixed sized cliques and if such a junction tree can be found efficiently [Ar85, ALS91]. Some of these problems are: INDEPENDENT SET, DOMINATING SET, GRAPH K-COLORABILITY, HAMILTONIAN CIRCUIT and CON-CRAINT SATISFACTION PROBLEMS [DP89].

Robertson and Seymour [RS95], among other key results, were the first to present an algorithm that finds a junction tree of a given graph $G$ such that its maximal clique size is at most a constant factor off optimal (They actually used a slightly different concept termed *branchwidth*). Reed [Re92] presents Robertson and Seymour's algorithm in a more accessible form and shows that its output is always less than 4 times the cliquewidth and the complexity is $O(k^2 3^{3k} n^2)$. Reed also gives an algorithm that errs by a factor of 5 and has a complexity $O(k^2 3^{4k} n \log n)$. Lagergren [La96] presents efficient parallel algorithms for this problem.

We offer an algorithm that finds a junction tree such that its largest clique is at most $(2\alpha + 1)$ times the cliquewidth where $\alpha$ is the approximation ratio for any approximation algorithm for the 3-way vertex cut problem. When using a $\frac{4}{3}$-approximation algorithm for the 3-way vertex cut problem ($\alpha = \frac{4}{3}$) due to [GVY94], our algorithm's complexity is $O(2^{4.66k} n \cdot poly(n))$ and it errs by a factor of 3.66 where $poly(n)$ is the running time of linear programming.

When $k = O(\log n)$, our algorithm, like previous ones, is polynomial. Consequently, it yields a polynomial inference algorithm for the class of Bayesian networks that have a logarithmic cliquewidth. Of course, one does not know a priori what is the cliquewidth of a given network and so a user must terminate the algorithm if the running time is too long, in which case, however, the running time of exact inference must be quite large as well. We show that for the class of Bayesian networks having a slightly larger than logarithmic cliquewidth, there exists no polynomial inference algorithm unless all NP-complete problems are solvable in less than exponential time.

In Section 3, we describe the algorithm and prove its performance guarantee. This algorithm is made as simple as possible to facilitate the proof. In Section 4, we describe several heuristics that improve the algorithm's average case performance. In Section 5, we describe the changes needed so that the algorithm takes into account vertices with different state space sizes. The modified algorithm guarantees that the logarithm of the size of the state space of the heaviest clique in the junction tree found is less than a constant factor off the optimal value. In Section 6 we describe experiments made using the graph Medianus I. In most instances our algorithm was superior to an enhanced greedy algorithm both in terms of the largest state space and in terms of the total state space. In Section 7 we discuss the extend to which our results can be improved.

## 2   The Junction Tree Algorithm

The junction tree algorithm is currently the most practical inference method for Bayesian networks. In this section we provide the relevant highlights of the junction tree algorithm. For details, consult [LS88, JLO90].

**Definition** A directed acyclic graph (DAG) is a graph with no directed cycles. In a DAG, $pa(v)$ denotes the set of parents of a vertex $v$. A *Bayesian network* is a DAG such that with each vertex $v$ we associate a finite set $D(v)$ called the state space of $v$ and a probability distribution $P(v|pa(v))$. The joint distribution of $V$ is given by $P(V) = \prod_{v \in V} P(v|pa(v))$.

The *updating problem* is to compute the posterior probability of a certain vertex given specific values to a set of other vertices.

The junction tree algorithm solves the updating problem as follows. For every vertex $v$, it connects every pair of $v$'s parents and removes the direction of all edges in the graph. The resulting graph is undirected (called the moral graph). Then, the moral graph is triangulated; edges are added until every cycle of length more than three has a chord. These are called *fill-in edges*. Once the graph is triangulated (or chordal), a tree of cliques, called the *junction tree*, is constructed. The junction tree algorithm then loads all probabilities into the junction tree and performs the calculations on the new structure.

**Definition** Let $G = (V, E)$ be a chordal graph. A *junction tree* of $G$ is a tree $\mathcal{H}$ such that each maximal clique $C$ of $G$ is a node in $\mathcal{H}$, and for every vertex $v$ of $G$, if we remove from $\mathcal{H}$ all nodes not containing $v$, the remaining (hyper) graph stays connected.

The single most important step of this algorithm is triangulation. There are many ways to add edges to a given graph until it becomes chordal. In particular, one can simply make a single large clique. However, the time for loading the probabilities and performing the calculations is proportional to the total state space given by $\sum_{C \in \mathcal{H}} \prod_{v \in C} |D(v)|$, which is dominated by the size of the maximal clique if all vertices have the same state space size. For example, if a maximal clique contains $m$ vertices and if their state space is of size two, then the probability table for this clique is of size $2^m$. The objective of triangulation is to find a triangulation such that the maximal clique size is as small as possible. Sections 3 and 4 are doing just that. In section 5, we describe the changes needed in order to account for varying state space sizes.

## 3   The Triangulation Algorithm

A natural approach to triangulate a graph $G = (V, E)$ is to use a divide and conquer technique. In each iteration a minimum set of vertices $X$ is found which removal from $G$ splits $G$ into two disconnected components having vertex sets $A$ and $B$ such that $A \cup B \cup X = V$. The set $X$ is called a *minimum vertex cut*. The algorithm proceeds on the two smaller problems $G[X \cup A]$ and $G[X \cup B]$, the subgraphs induced from $G$ by the vertex sets $X \cup A$ and $X \cup B$ respectively.

Each subgraph is triangulated such that $X$ becomes a clique in it.

While this approach yields a triangulated graph, the size of the cliques produced may grow up to an $O(n)$ factor off their initial size if in each step one of the graphs shrinks only by a constant number of vertices and the vertex cut found in each step has many edges connecting it to previously found cuts. Robertson and Seymour, Reed, and Kloks all provide clever modifications that prevent the initial clique $X$ from growing beyond a constant factor off its initial size.

We provide an algorithm that is similar to previous ones except that rather than dividing the graph to two subproblems, we divide it to three subproblems. As a procedure, we use an $\alpha$-approximation algorithm for the 3-way vertex cut problem. The 3-way vertex cut problem is defined as follows: given a weighted undirected graph and three vertices, find a set of vertices of minimum weight whose removal leaves each of the three vertices disconnected from the other two. This problem is known to be NP-hard [Cu91]. There exists a simple 2-approximation algorithm, that is, the weight of its output is no more than twice the weight of an optimal 3-way vertex cut. A polynomial $\frac{4}{3}$-approximation algorithm for the 3-way vertex cut problem is reported in [GVY94] (Actually, their algorithm is a $(2 - \frac{2}{k})$-approximation algorithm that finds k-way vertex cuts).

Our algorithm produces a triangulated graph whose maximal clique size is less than $(2\alpha+1)k$ where $k$ is the cliquewidth of $G$ and $\alpha$ is the ratio between the weight of the 3-way vertex cut found by the algorithm we use and the optimal 3-way vertex cut. For $\alpha = \frac{4}{3}$, obtained by using Garg et al's algorithm, our approximation algorithm yields a triangulation having a cliquewidth bounded by $3\frac{2}{3}k$.

**Definition** Let $G = (V, E)$ be a graph. A *decomposition* of $G$ is a partition $(X, A, B, C)$ of $V$, where $A$ and $B$ are non-empty sets, such that there are no edges between $A, B$ and $C$.

**Definition** Given an integer $k \geq 1$, a real number $\alpha \geq 1$, a graph $G = (V, E)$ such that $|V| \geq (2\alpha+1)k$, and a subset of vertices $W \subseteq V$, a decomposition $(X, A, B, C)$ of $G$ is called a *W-decomposition* wrt $(k, \alpha)$ if $|W| < (\alpha+1)k$, $|X| \leq \alpha k$, $|(W \cap A) \cup X| < (\alpha+1)k$, $|(W \cap B) \cup X| < (\alpha+1)k$, and $|(W \cap C) \cup X| < (\alpha+1)k$.

For example, suppose $G$ is the chain $a - b - c - d - e$. The triplet $X = \{c\}, A = \{a, b\}, B = \{d, e\}$ and $C = \emptyset$ is a decomposition of $G$. Given $W = \{b, d\}$, this decomposition is a W-decomposition of $G$ wrt $k = 1$ and $\alpha = 2$. Given $W = \{b, c\}$, the triplet $X = \{d\}, A = \{a, b, c\}$, $B = \{e\}$ and $C = \emptyset$ is not a W-decomposition of $G$ wrt $(k = 1, \alpha = 2)$ because $|(W \cap A) \cup X| = 3$.

The triangulation algorithm is given in Figure 1. In

## ALGORITHM Triangulate(G,W,k)

**Input:**   *An undirected graph $G(V, E), W \subseteq V, k$.*
**Output:**   *A triangulation of $G$ such that $W$ is made a clique and such that the size of the largest clique $< (2\alpha + 1)k$ (Success) or, a valid statement that the cliquewidth of $G$ is larger than $k$ (Failure).*

**If** $|V| < (2\alpha + 1)k$ **then** make a clique out of $G$
**else**
   Find a W-decomposition
                           $(X, A, B, C)$ of $G$ wrt $(k, \alpha)$;
   **If** not found return "cliquewidth $> k$"
   $W_A \leftarrow W \cap A, W_B \leftarrow W \cap B, W_C \leftarrow W \cap C$;
   **call** Triangulate($G[A \cup X], W_A \cup X, k$);
   **call** Triangulate($G[B \cup X], W_B \cup X, k$);
   **call** Triangulate($G[C \cup X], W_C \cup X, k$);
   make a clique of $G[W \cup X]$.

Figure 1: The triangulation algorithm

order to triangulate a graph $G$ having a cliquewidth $k$ we call *Triangulate* $(G, \emptyset, k)$. When the algorithm is called with $W = \emptyset$, the size of the second argument of *Triangulate* in each recursive call is (strictly) less than $(\alpha+1)k$ because, when $|V| \geq (2\alpha+1)k$, by definition of W-decompositions, the sets $W_A \cup X, W_B \cup X, W_C \cup X$ which are the arguments passed in the recursive calls, contain less than $(\alpha + 1)k$ vertices, respectively. Figure 2 shows a graph and how it splits into three subgraphs in a recursive call of *Triangulate* . The set $W$ serves to monitor the shrinking rate of the size of the subproblems in each recursive call.

The next two lemmas show that a W-decomposition must exist or the cliquewidth is greater than $k$, in which case the algorithm outputs correctly this fact. Consequently, a naive way to use this algorithm is to repeatedly call TRIANGULATE($G, \emptyset, k$) starting with $k = 1$ and incrementing $k$ by 1 whenever the algorithm fails to triangulate $G$. In the next section, we provide implementation details and a complexity analysis.

**Lemma 1** *Given a graph $G(V, E)$ with a cliquewidth $\leq k$, $|V| \geq k+2$, and a subset of vertices $W, |W| > 1$, there exists a decomposition $(X, A, B, C)$ of $G$ such that $|X| \leq k$, $|W \cap A| \leq \frac{1}{2}|W|$, $|W \cap B| \leq \frac{1}{2}|W|$ and $|W \cap C| \leq \frac{1}{2}|W|$.*

**Proof:** A constructive proof of similar claims is given in [Kl94, Lemma 2.2.9]. Let $H(G)$ be a junction tree of $G$ with a cliquewidth $\leq k$. Add edges until all cliques in this junction tree become of size $k$. Call the resulting junction tree $T(G)$. Now consider the following algorithm. Start with any clique $X$ in $T(G)$. If there is no connected component in $G[V \setminus X]$ which has more than $\frac{1}{2}|W|$ vertices of $W$, then stop. Otherwise, let $S$ be a component in $G[V \setminus X]$ which has more than $\frac{1}{2}|W|$ vertices of $W$. There exists a vertex $y$ in $S$ which has $k - 1$ neighbors in $X$ in the graph $T(G)$ (viewed as a chordal graph). Let $x$ be the vertex in $X$ that is not a neighbor of $y$ in $T(G)$. Define $Y = X \setminus \{x\} \cup \{y\}$. Note that $Y$ also has $k$ vertices. The algorithm continues with $Y$.

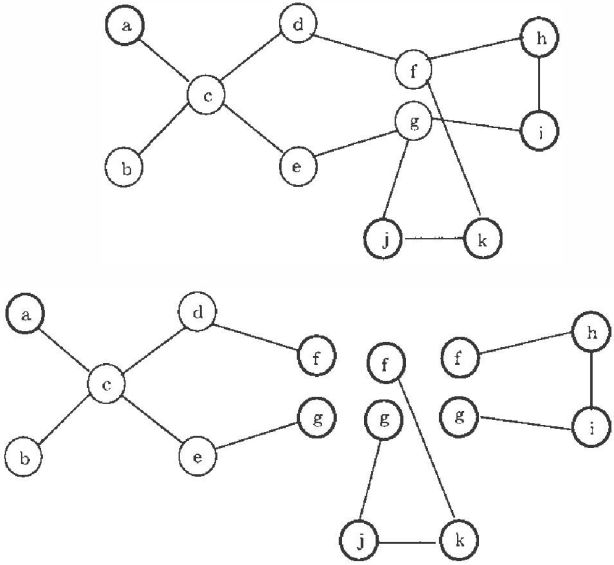

Figure 2: An example of one level of a recursive call with $k = 3$ and $\alpha = 1$. Highlighted vertices are in $W$ and $X = \{f, g\}$. The three graphs at the bottom are passed as arguments to the next level of recursion.

To show that this algorithm terminates, we prove that in each step of the algorithm one of two conditions is met. Hereafter, the component which includes the largest part of $W$ will be called the *main* component. The first condition is that the number of vertices in the main component decreases and the number of vertices of $W$ in the main component does not increase. The second condition is that the number of vertices of $W$ in the main component decreases.

Notice that $G[V \setminus Y]$ has two types of components. One type consist only of vertices in $S \setminus \{y\}$. If the main component of $G[V \setminus Y]$ is among these, the number of vertices is decreased and the number of vertices of $W$ does not increase. The other type of components consist only of vertices of $\{x\} \cup V \setminus (S \cup X)$. The total number of vertices of $W$ in this set is less than $\frac{1}{2}|W|$ because $S$ contains more than half the vertices of $W$. Hence, in this case, the number of vertices from $W$ in the main component decreases by at least one. Consequently, the algorithm terminates.

Suppose now that $X$ is the final clique considered by this algorithm. If $G[V \setminus X]$ has two or more non empty components, then group them into three sets to form the desired decomposition. Otherwise, there is only one component in $G[V \setminus X]$. Consequently, the clique $X$ is a leaf in the junction tree $T(G)$. Since $|V|$ contains at least $k + 2$ vertices, and there is only one component in $G[V \setminus X]$, there exists a unique clique $Y$ that contains $k - 1$ vertices of $X$ and which is not a leaf in $T(G)$. The graph $G[V \setminus Y]$ has at least two connected components and each contains less than half the vertices of $W$ (because $|W| > 1$). $\square$

**Lemma 2** *Given an integer $k \geq 1$, a real number $\alpha \geq 1$, a graph $G(V, E)$ with $|V| \geq (2\alpha + 1)k$ and a subset of vertices $W$ such that $|W| < (\alpha + 1)k$, there exists a $W$-decomposition $(X, A, B, C)$ of $G$ wrt $(k, \alpha)$ or the cliquewidth of $G$ is larger than $k$.*

**Proof:** Let $G$ be a graph with a cliquewidth $\leq k$. If $|W| \leq 1$, then let $X$ be any minimal vertex cut that does not contain a vertex of $W$. If $|X| \leq k$, the resulting decomposition is a W-decomposition wrt $(k, \alpha)$. Otherwise, the cliquewidth is larger than $k$.

Suppose $|W| > 1$. Let $(X, A, B, C)$ be a decomposition of $G$ with the properties guaranteed by Lemma 1. We will prove that $(X, A, B, C)$ is also a W-decomposition wrt $(k, \alpha)$. If it were not, then assume that $|(W \cap A) \cup X| \geq (\alpha + 1)k$, this inequality implies that $|W \cap A| \geq \alpha k$ because $|X| \leq k$. But according to Lemma 1, we have $|W| \geq 2|W \cap A|$. Consequently, $|W| \geq 2\alpha k$ in contradiction to its given size which is smaller than $(\alpha + 1)k$. Hence, if the cliquewidth of $G \leq k$, then there is a W-decomposition wrt $(k, \alpha)$. Equivalently, if there isn't a W-decomposition wrt $(k, \alpha)$, the cliquewidth must be larger than $k$. $\square$

**Theorem 3** *If $G(V, E)$ is a graph with $n$ vertices, $k \geq 1$ an integer, $\alpha \geq 1$ a real number, and $W$ is a subset of $V$ such that $|W| < (\alpha + 1)k$, then Triangulate$(G, W, k)$ triangulates $G$ such that the vertices of $W$ form a clique and such that the size of a largest clique of the triangulated graph $< (2\alpha + 1)k$ or the algorithm correctly outputs that the cliquewidth of $G$ is larger than $k$.*

**Proof:** If the algorithm outputs that the cliquewidth of $G$ is larger than $k$, then this is a valid statement by lemma 2. Assume the algorithm does not produce this output.

The algorithm always terminates because in every recursive call of *Triangulate* the graphs $G[A \cup X]$, $G[B \cup X]$ and $G[C \cup X]$ have less vertices than $G[A \cup B \cup C \cup X]$ since $A$ and $B$ are not empty.

Next, we show that the algorithm returns a triangulated graph. We prove this by induction using the recursive structure of the algorithm. Clearly the claim is true if $|V| < (2\alpha + 1)k$. Assume $|V| \geq (2\alpha + 1)k$. By induction the recursive call *Triangulate* $(G[A \cup X], W_A \cup X, k)$ returns a triangulation of $G[A \cup X]$, such that $W_A \cup X$ is a clique. Similarly, for $B$ and $C$. The algorithm then makes a clique of $W \cup X$. Consequently, the graphs $G[A \cup W \cup X]$, $G[B \cup W \cup X]$ and $G[C \cup W \cup X]$ are triangulated as well. Since the intersection of these triangulated graphs is a clique, the union must also be triangulated.

It remains to show that the cliquewidth of the triangulated graph is less than $(2\alpha + 1)k$. This is clearly true if $|V| < (2\alpha+1)k$. Hence assume $|V| \geq (2\alpha+1)k$. Let $M$ be a largest clique in the triangulated graph. There are two cases to consider. If $M$ contains no vertex of $A \setminus W$, $B \setminus W$ and $C \setminus W$, then $M$ contains only vertices of $W \cup X$. Consequently, $|M| = |W \cup X| \leq |W| + |X| < (\alpha+1)k + \alpha k$, and the cliquewidth is less than $(2\alpha+1)k$ as claimed. If $M$ contains a vertex of $A \setminus W$, then it contains no vertex of $B \cup C$ because there are no edges between $A$ and $B \cup C$. Hence $M$ is a clique in the triangulation of $G[A \cup X]$. By induction we know that $|M| < (2\alpha+1)k$. □

Note that Lemma 2 and Theorem 3 hold for every $\alpha \geq 1$. However, in order to find a W-decomposition wrt $(k, \alpha)$ sufficiently fast (Lemma 2 only guarantees existence), we choose $\alpha$ to be the approximation factor of an algorithm for the 3-way vertex cut problem, an algorithm which we employ for finding a W-decomposition. We now give an algorithm that finds a W-decomposition wrt $(k, \alpha)$ where $\alpha$ is chosen as just described. Then we will argue for correctness.

For every possible selection of four disjoint subsets $W_A, W_B, W_C, W_X$ of $W$, such that $|W_A| \geq |W_B| \geq |W_C|$, we show how to check if there exists a W-decomposition $(X, A, B, C)$ wrt $(k, \alpha)$, such that $W_A \subseteq A, W_B \subseteq B, W_C \subseteq C$ and $W_X \subseteq X$. There are at most $4^{|W|}$ choices to divide $W$ into four set of vertices $W_A, W_B, W_C, W_X$.

Let $W_A, W_B, W_C, W_X$ be a particular selection. We consider two cases, 1) $|W_A| < k$ and 2) $|W_A| \geq k$. Each case uses a different procedure.

Procedure I ($|W_A| < k$): Remove $W_X$ from the graph, add three dummy vertices $v_a, v_b$ and $v_c$ each connected to all the vertices in $W_A$, $W_B$ and $W_C$, respectively. Set the capacity of all vertices in $W_A \cup W_B \cup W_C$ to infinity and the capacity of all other vertices to one. Find an $\alpha$-approximation 3-way vertex cut $Y$ which splits $v_a$, $v_b$ and $v_c$ into three disconnected components. Note that, due to the capacities selected, it must split $W_A, W_B$ and $W_C$ to three disconnected components as well. Now let $X$ be $Y \cup W_X$, $A$ be the union of the connected components of $G[V \setminus X]$ such that $W_A \subseteq A$, $B$ be the union of the connected components of $G[V \setminus X]$ such that $W_B \subseteq B$, and $C = V \setminus (A \cup B \cup X)$. If $|X| < (\alpha+1)k - |W_A|$ and $|X| \leq \alpha k$ then output $(X, A, B, C)$ as the desired W-decomposition wrt $(k, \alpha)$ (because $|W_A| \geq |W_B| \geq |W_C|$).

Procedure II ($|W_A| \geq k$): Remove $W_X$ from the graph, add a dummy vertex $v_a$ that is connected to all the vertices in $W_A$, and add another dummy vertex $v_{bc}$ that is connected to all vertices in $W_B$ and $W_C$. Set the capacity of all vertices in $W_A \cup W_B \cup W_C$ to infinity and the capacity of all other vertices to one. Find a minimum vertex cut $Y$ which splits $v_a$ and $v_{bc}$ into two disconnected components. Note that it must split $W_A$ and $W_B \cup W_C$ as well. Finding a minimum vertex cut is done by any of the well known max-flow/min-cut algorithms. Now let $X$ be $Y \cup W_X$, $A$ be the union of the connected components of $G[V \setminus X]$ such that $W_A \subseteq A$, $B = V \setminus (A \cup X)$, and $C = \emptyset$. If

$|X| < (\alpha+1)k - |W_A|$, $|X| < (\alpha+1)k - |W_B \cup W_C|$ and $|X| \leq \alpha k$ then output $(X, A, B, C)$ as the desired W-decomposition wrt $(k, \alpha)$.

Now we will show that if a W-decomposition wrt $(k, \alpha)$ exists, as guaranteed by Lemma 2, then either procedure I or procedure II will find a W-decomposition wrt $(k, \alpha)$ for some choice of $W_A, W_B, W_C, W_X$. Let $(X', A', B', C')$ be a decomposition of $G$ with the properties guaranteed by Lemma 1. Let $W_A = A' \cap W$, $W_B = B' \cap W, W_C = C' \cap W$ and $W_X = X' \cap W$, and assume without loss of generality that $|W_A| \geq |W_B| \geq |W_C|$. Procedure I for $|W_A| < k$, and procedure II for $|W_A| \geq k$ both generate for this choice of $W_A, W_B, W_C, W_X$, a decomposition $(X, A, B, C)$. We now show that in either case this decomposition is a W-decomposition wrt $(k, \alpha)$.

Case 1: $|W_A| < k$. The set of vertices $X' \setminus W_X$ is a 3-way vertex cut for the sets $W_A$, $W_B$, and $W_C$ in the graph $G[V \setminus W_X]$. An $\alpha$-approximation algorithm for the 3-way vertex cut problem outputs a cut $Y$, such that $|Y| \leq \alpha |X' \setminus W_X|$. Since $\alpha \geq 1$, we get $|Y \cup W_X| \leq \alpha |X'|$. Consequently, $|Y \cup W_X| \leq \alpha k$ because $|X'| \leq k$. Finally, $|W_A \cup (Y \cup W_X)| < k + \alpha k = (\alpha+1)k$. Thus all the conditions for a W-decomposition wrt $(k, \alpha)$ are met.

Case 2: $|W_A| \geq k$. The set of vertices $X' \setminus W_X$ is a vertex cut for the sets $W_A$, $W_B \cup W_C$ in $G[V \setminus W_X]$. A max flow/min-cut algorithm outputs a vertex cut $Y$ such that $|Y| \leq |X' \setminus W_X|$. Consequently $|Y \cup W_X| \leq k$ because $|X'| \leq k$. Finally, since $|W| \geq 2|W_A|$ (Lemma 1), we get $|W_A \cup (Y \cup W_X)| < \frac{\alpha+1}{2}k + k \leq (\alpha+1)k$. Hence from $|W_B \cup W_C| < \alpha k$ it follows that $|(W_B \cup W_C) \cup (Y \cup W_X)| < \alpha k + k = (\alpha+1)k$. Thus all the conditions for a W-decomposition wrt $(k, \alpha)$ are met.

## 4    Implementation and Complexity

The algorithm presented in the previous section can be improved substantially by three adjustments: processing the input of the algorithm, changing the termination condition of the recursion, and processing the output of the algorithm. We shall first describe these changes and demonstrate the algorithm on a simple example. Then, we provide further implementation details and analyze the algorithm's complexity.

The input graph of the algorithm may contain vertices such that all their neighbors are connected. A vertex $v$ is called *simplicial* in $G$ if its neighbors $N(v)$ form a clique. Before calling *Triangulate*, starting with the input graph $G$, we repeatedly remove every simplicial vertex from the current graph. The resulting graph $G'$ has a cliquewidth no larger than that of $G$, and if $G'$ is triangulated, then $G$ is triangulated as well. Hence, this preprocessing step retains the validity of the algorithm. This step improves the running time complexity whenever simplicial vertices are found.

The termination condition of the recursion is that whenever $|V| < (2\alpha + 1)k$ a clique is formed out of $G$. However, instead of a clique, it suffices to produce a junction tree of $G$ in which $W$ is a clique. This step is done by forming a clique of $W$ and then completing it to a junction tree of $G$ by any of the known greedy heuristics. The proof of Theorem 3 remains valid without any change. Consequently, the worst case approximation is not affected. However, in many instances the approximation is improved.

The output of the algorithm is a triangulated graph $T(G)$ which is not necessarily minimal. This means that some edges that were added (fill-in edges) might possibly be removed and the resulting graph remains triangulated. Kjaerulff provides an algorithm that, given a triangulation of a graph $G$ and an ordering on its vertices, produces a minimal triangulated graph [Kj90]. We use Kjaerulff's algorithm with an ordering that is determined as follows. First in the ordering are the simplicial vertices in the order in which they are removed from $G$. The order of the remaining vertices is determined recursively while running $Triangulate$; In each level of the recursion, the vertices in $X \setminus W$ follow those in $A \setminus W$, those in $B \setminus W$ and those in $C \setminus W$.

We now demonstrate the effects of these modifications on the graph depicted in Figure 2 (assuming $W = \emptyset$). If simplicial vertices are removed, then the remaining graph does not contain the vertices $a$ and $b$. The next phase, when $k = 3$ and $\alpha = 1$, creates three cliques: $\{c, d, e, f, g\}$, $\{f, g, h, i\}$ and $\{f, g, j, k\}$, in addition to $\{a, c\}$ and $\{b, c\}$ due to the simplicial vertices. Finally, applying Kjaerulff's minimization algorithm removes the edges $(f, i), (f, j), (c, f), (c, g), (d, g)$ yielding an optimal junction tree.

The time complexity of running $Triangulate$ with a given $k$ is the time it takes to find a W-decomposition times the number of nodes in the recursion tree which is at most $n$. The time it takes to verify whether a choice $W_A, W_B, W_C, W_X$ can generate a W-decomposition with respect to $(k, \alpha = \frac{4}{3})$ takes $poly(n)$ which is the time it takes to run Garg et al's $\frac{4}{3}$-approximation algorithm for the 3-way vertex cut problem. This polynom is quite high as it is the complexity linear programming. (A more practical algorithm, without a complexity guarantee, is the simplex algorithm). Thus the complexity of running $Triangulate$ with a given $k$ is $O(4^{(1+\alpha)k}n \cdot poly(n))$ where $\alpha = \frac{4}{3}$, because there are at most $4^{|W|}$ choices and $|W| < (\alpha + 1)k$. Since, in the worst case, the algorithm is run for $i = 1$ up to the cliquewidth of $G$, the total running time is $O(\sum_{i=1}^{k} 4^{2.33i}n \cdot poly(n))$ which is $O(2^{4.66k}n \cdot poly(n))$. The size of the largest clique in the output is at most $2\alpha + 1 = 3.66$ times the cliquewidth.

In a simpler implementation we use a straightforward 2-approximation algorithm for finding a 3-way vertex cut; Find a minimum $a - b$ vertex cut between $v_a$ and $v_b$, a minimum vertex cut $a - c$ between $v_a$ and $v_c$ and a minimum vertex cut $b - c$ between $v_b$ and $v_c$. The output vertex cut is the union of any two of the three vertex cuts. This output is clearly a 3-way cut and it is at most twice the optimal weight because each of the three cuts weighs less than an optimal 3-way vertex cut. Finding each vertex cut is done using a max flow/min-cut algorithm which takes $O(kn^2 \log n)$. This algorithm for the 3-way vertex cut is analogous to the one described in [DJPSY92] for the edge multiway cut. A more clever implementation using Reed's arguments can find an appropriate vertex cut in $O(k^2n)$. Consequently, since $\alpha = 2$, the total complexity is $O(k^2 4^{3k}n^2)$ and the largest clique in the output is at most 5 times the cliquewidth.

In practice, our algorithm encountered complexity is substantially less. The set $W$ is almost always less than $(1 + \alpha)k$ and in most cases it is less then $k$ which implies that the complexity encountered is proportional to $2^{2k}$ rather than to $2^{4.66k}$. Furthermore, when a W-decomposition $(X, A, B, C)$ exists, it is often the case that $W$ consists of two subsets and the third is empty, in which case the algorithm for finding a 3-way vertex cut is not activated (as is the case in the graph of Section 6). In addition, instead of increasing $k$ by one whenever $Triangulate$ fails on the input $k$, we can increase it to the minimal value $k^*$ for which a decomposition that was tested wrt $(k, \alpha)$ was found to be a W-decomposition wrt $(k^*, \alpha)$ $(k^* > k)$.

Finally, note that Theorem 3 provides only a worst case bound of $2\alpha + 1$ for the ratio between the size of the largest clique and the cliquewidth of the given graph. However, if for an integer $k$, $Triangulate$ produces a triangulation having a largest clique of size $l$ and the algorithm fails for $k - 1$ (it is run for $i = 1..k$ until in succeeds), then the ratio $l/k$ is an upper bound for the ratio between the output and the cliquewidth of $G$ because the cliquewidth must be greater than $k - 1$. This bound is much tighter than $2\alpha + 1$ because it takes into account the given graph and the specific steps made by $Triangulate$. It is an instance-specific posteriori bound rather than a worst case a priori bound. The bound $l/k$ is produced by the algorithm in order to inform the user about the quality of the junction tree found.

## 5    The Weighted Problem

It remains to describe the changes needed in order to account for different state spaces of each vertex. The weight $w(v)$ of a vertex $v$ is the logarithm (base 2) of its state space size and the weight of a clique is the sum of the weights of its constituent vertices. Note that the weight of a vertex with a binary state space is 1 and the weight of other vertices is larger than 1. Our optimality criterion is now the weighted cliquewidth of $G$. The *weighted cliquewidth* of $G$ is the weight of the heaviest clique in a junction tree of $G$ in which the weight of the heaviest clique is minimized.

To minimize the heaviest clique, we modify the algo-rithm as follows. We find a *weighted W-decomposition* wrt $(m, \alpha)$, namely, a decomposition $(X, A, B, C)$ of $G = (V, E)$, where $w(V) \geq (2\alpha + 1)m$, such that $w(W) < (\alpha + 1)m$, $w(X) \leq \alpha m$, $w((W \cap A) \cup X) < (\alpha + 1)m$, $w((W \cap B) \cup X) < (\alpha + 1)m$ and $w((W \cap C) \cup X) < (\alpha + 1)m$. Once the termination condition is met, namely, $w(V) < (2\alpha + 1)m$, we apply the following greedy algorithm which is called the *minimum weight heuristics*: repeatedly, select a vertex $v$ which forms with its neighbors $N(v)$ a set of minimum weight, remove it from the current graph, and make $N(v)$ a clique. We call this modified algorithm *W-Triangulate*.

The following claim holds.

**Theorem 4** *If $G$ is a graph with $n$ vertices, $m$ and $\alpha \geq 1$ are real numbers, and $W$ is a subset of $V$ such that $w(W) < (\alpha + 1)m$, then W-Triangulate$(G, W, m)$ triangulates $G$ such that the vertices of $W$ form a clique and such that the weight of a heaviest clique of the triangulated graph $< (2\alpha + 1)m$ or the algorithm correctly outputs that the weighted cliquewidth of $G$ is larger than $m$.*

**Proof:** Theorem 3 and Lemmas 1 and 2 remain valid when the cardinality of sets is replaced with their weight and $k$ is replaced with $m$. □

Theorem 4 states that in the junction tree found by *W-Triangulate* the weight of the heaviest clique is less then $2\alpha + 1$ times the weighted cliquewidth.

The complexity of *W-Triangulate* depends on the maximum size of $W$ throughout the recursive calls which we denote by $s$. The complexity of *W-Triangulate* is $O(k^2 4^{3s} n^2)$ when the 2-approximation algorithm for the 3-way vertex cut problem is used. The heaviest clique in the resulting junction tree is at most 5 times the weighted cliquewidth. Since, $k \leq s \leq \min\{m, n\}$, this complexity is comparable to the complexity of inference on the resulting junction tree which is $O(2^{5m} n)$ and it is smaller than the complexity of inference if state spaces are sufficiently large. Usually $s$ is closer to the cliquewidth $k$ than to $m$ or $n$.

Indeed, a more subtle modification of *Triangulate* yields an algorithm that is exponential in the cliquewidth $k$ rather than in $s$.

**Theorem 5** *Let $G$ be a graph with $n$ vertices having a weighted cliquewidth $m$ and a cliquewidth $k$. Then, there exists an algorithm $W^*$-Triangulate having a complexity of $O(c^k n^a)$, where $a$ and $c$ are constants, which finds a junction tree such that the weight of its heaviest clique is at most a constant factor off $m$.*

The algorithm $W^*$-Triangulate gets two parameters $k$ and $m$. In each step, it finds a decomposition $X$ which is bounded both by $(2\alpha + 2)m$ and by $(2\alpha + 2)k$. Thus, it guarantees that the weight and, simultaneously, the size of $W$ will not grow too much. This algorithm cannot outperform *W-Triangulate* (in the experiments reported herein) because in all our experiments whenever the weight of $W$ was small, the size was small as well.

|   | W-Triangulate | | | Eq | Greedy | | |
|---|---|---|---|---|---|---|---|
|   | # | $\Delta_{ave}$ | $\Delta_{max}$ | # | # | $\Delta_{ave}$ | $\Delta_{max}$ |
| M | 75 | .62 | 2.35 | 16 | 9 | .3 | .93 |
| T | 73 | .64 | 2.37 | 11 | 16 | .42 | 1.22 |

Figure 3: The results for 100 runs on Medianus I. The first line records the differences on the average and in the extreme case of the logarithm of the heaviest clique. In 16 instances the algorithms produced equal output. The second line records the same information regarding the logarithm of the total state space.

## 6    Experimental Results

Kjaerulff (1990) has tested several heuristic algorithms for constructing junction trees for two graphs that were used for a medical application: Medianus I and Medianus II. His experiments show that the minimum weight heuristics enhanced by removing redundant fill-in edges is superior to all other heuristics that were considered. We will compare *W-Triangulate* with this enhanced minimum weight heuristics on Medianus I. This graph contains 43 vertices and 110 edges. We use two optimality criteria for the comparison, the logarithm of the state space size of the heaviest clique denoted by $M$ and the logarithm of the total state space denoted by T. Criterion $M$ is the one that served to develop *W-Triangulate* and $T$ is the one that optimizes the construction of the probability tables for the resulting junction tree.

The two algorithms were run on Medianus I with state sizes randomly selected from the range 3 to 21 with an average of approximately 6 (as in [Kj90]). One hundred random runs were made. In 68 runs our algorithm has outperformed the enhanced minimum weight heuristics in both optimality criteria. Figure 3 shows that the averaged improvement of $T$ was 0.64 and the maximum improvement was 2.37 which amounts to a reduction of storage by a factor of about 5. In the 16 instances in which the greedy method was more successful, the difference was at most 1.22. Of course, to obtain the best results one can simply run both algorithms.

When the state space size of each vertex was selected between 6 and 32 with an average of 13, we found two graphs in which $T$ is approximately 30 using *W-Triangulate* and $T$ is approximately 34.5 using the enhanced greedy algorithm which implies that instead of 1GB of memory which we need for storing the conditional probabilities, the greedy algorithm would have used over 20GB. In general, as the state spaces increase our algorithm becomes far better than the enhanced minimum weight heuristics.

Recall that the algorithm *W-Triangulate(G, ∅, k)* is run with increasing values of $k$ until a triangulation is found. We have recorded the number of vertices $l$ in the largest clique (in size) of the junction tree found by *W-Triangulate(G, ∅, k)* when it succeeded and compared it to the value of $k$. Let $\Delta = l - k$. The maximal clique size found is of size $l$ while the cliquewidth is larger than $k - 1$ (because the algorithm failed with $k - 1$ as an input). Then, $\Delta$ was 0 in 6 graphs (provably optimum size), 1 in 14 graphs (at most one vertex off optimum), 2 in 29 graphs, 3 in 48 graphs and 4 in 3 graphs. The worst case upper bound on the ratio between the size of the largest clique and the unknown cliquewidth was $l/k = 10/6$ rather than 3.66 which is guaranteed in theory. Indeed, one cannot hope to improve the junction tree too much on this graph.

We also collected some statistics on the running time complexity. We counted the number of partitions made each time a W-decomposition is constructed. The count for Medianus I was always far less than $4^k$ rather than $4^{3k}$ which is the worst case bound. The recursion depth was at most 3. The algorithm runs in less than a minute for most graph instances but occasionally it takes up to two minutes. On these examples Robertson and Seymour's algorithm runs faster and obtains identical results.

## 7    Discussion

We presented an algorithms that finds a junction tree in which the largest clique is no more than 3.66 times the cliquewidth. If the cliquewidth of $G$ is of size $k = O(\log n)$, then our approximation algorithm is polynomial since its complexity is $O(2^{4.66k}n \cdot \text{poly}(n))$ where $poly(n)$ is the complexity of linear programming. It is well known that inference in an optimal junction tree with binary variables takes $O(2^k n)$ which is polynomial for a logarithmic cliquewidth. Thus, inference done using the junction tree produced by our algorithm, as well as by Robertson and Seymour's algorithm, is guaranteed to be polynomial as well because if we err at most by a constant factor, the time of inference is at most the optimal time raised to some power and so inference stays polynomial. Note that the heuristic constructions of junction trees which do not guarantee a constant error bound are not polynomial.

The claim that finding the cliquewidth of a graph is polynomial if $k = O(\log n)$ means that for every sequence of graphs $G_{n,k_n}, n = 1, \ldots,$ with $n$ vertices and a cliquewidth $k_n$, our algorithm finds the cliquewidth in polynomial time if $k_n < c \log n$ for $n \geq n_0$ where $n_0$ and $c$ are constants.

The natural question to raise is whether a polynomial inference algorithm exists if the cliquewidth grows a bit faster than logarithmic, say $k_n = O(\log^{1+\epsilon} n)$ for $\epsilon > 0$. We now show that if a polynomial inference algorithm exists for all networks having such a cliquewidth growth, then every inference problem can be solved in subexponential time which implies that every NP-complete problem can be solved in subexponential time due to Cooper's reduction from 3-SAT [Co90]. Let $G_{l,k_l}$ be a sequence of graphs for which the cliquewidth grows at a slightly faster rate than logarithmic. Suppose an inference problem is given on each network in this sequence. Examine the network in the sequence with $l$ vertices. Add isolated vertices to the given network. The cliquewidth remains unchanged and is at most $l$. When enough vertices are added (i.e., $l = O(\log^{1+\epsilon} n)$), we use the assumed polynomial inference algorithm to solve the inference problem of the augmented graph which also solves the original inference problem. The complexity of this assumed algorithm is polynomial in $n$—the number of vertices with the added isolated ones—which is subexponential in $l$. Consequently, this algorithm solves an arbitrary inference problem in less than exponential time (in $l$).

One must emphasize that this negative result means that probably there are some hard graphs for inference among those having a supper logarithmic cliquewidth. We believe that actually all such graphs are hard for inference if the proper conditional tables are used (e.g. polytrees can have an arbitrary large cliquewidth and they are still solvable for specific conditional tables, i.e., the noisy-or model [Pe88]).

Our results could possibly be improved in the following direction. One extension of our work is to construct an algorithm that finds an optimal junction tree wrt the weighted cliquewidth with a complexity of optimal inference, i.e., $O(2^k n)$, or errs by a factor smaller than 3.66. Our algorithm can yield at best a factor of 3 if an efficient exact algorithm is found for the 3-way vertex cut problem for graphs with bounded cliquewidth (The existence of such an algorithm is hinted parenthetically in [DJPSY92] but we have not yet pursued this direction).

As a final comment, let us shed light on the common utterance used by the AI community, that "inference is easy in sparse graphs". Recall that if the cliquewidth is of size $k$, then the graph has no more than $kn$ edges (see e.g., Section 4). Hence, sparse graphs in the context of easy inference should mean that the cliquewidth is of size $O(\log n)$, which allows a polynomial inference algorithm, and implies that there are no more than $O(n \log n)$ edges in the graph.

## Acknowledgment

We thank Seffi Naor for pointing us to the term treewidth, for helping us prove that a polynomial inference algorithm is not likely to exist if the cliquewidth is larger than logarithmic, and for pointing us to [GVY94]. We thank Leonid Zusin for showing us examples of poor junction trees produced by various greedy algorithms. We thank the reviewers for their helpful comments and for the references they provided us, in particular, Reed's paper.